\documentclass[aps,prb,groupedaddress,amsmath,eqsecnum,twocolumn]{revtex4}

\usepackage{graphicx}
\usepackage{dcolumn}
\usepackage{bm}
\usepackage{color}

\usepackage{amssymb}

\newcommand\beq{\begin{equation}}
\newcommand\eeq{\end{equation}}
\newcommand\beqa{\begin{eqnarray}}
\newcommand\eeqa{\end{eqnarray}}
\newcommand{\nn}{\nonumber\\}

\newcommand{\Tt}{T^\text{tr}}
\newcommand{\Tr}{T^\text{rot}}
\newcommand{\zt}{\xi^\text{tr}}
\newcommand{\zr}{\xi^\text{rot}}
\newcommand{\at}{\widetilde{\alpha}}
\newcommand{\bt}{\widetilde{\beta}}
\newcommand{\al}{\alpha}
\newcommand{\be}{\beta}
\newcommand{\rr}{\mathbf{r}}
\newcommand{\vv}{\mathbf{v}}
\newcommand{\cc}{\mathbf{v}}
\newcommand{\ww}{\bm{\omega}}
\newcommand{\q}{\kappa}
\newcommand{\gh}{\mathbf{v}_{12}}
\newcommand{\cca}{\mathbf{v}_1}
\newcommand{\ccb}{\mathbf{v}_2}
\newcommand{\ca}{{v}_1}
\newcommand{\cb}{{v}_2}
\newcommand{\wwa}{\bm{\omega}_1}
\newcommand{\wwb}{\bm{\omega}_2}
\newcommand{\wa}{{\omega}_1}
\newcommand{\wb}{{\omega}_2}
\newcommand{\kk}{\widehat{\bm{\sigma}}}
\newcommand{\wwwa}{\bm{\mathcal{V}}_{1}}
\newcommand{\wwwb}{\bm{\mathcal{V}}_{2}}
\newcommand{\wwwab}{\bm{\mathcal{V}}_{12}}
\newcommand{\da}{\sigma}
\newcommand{\x}{\times}
\newcommand{\SSab}{\mathbf{S}_{12}}
\newcommand{\JJ}{\bm{\Delta}_{12}}
\newcommand{\JJw}{\overline{\bm{\Delta}}_{12}}
\newcommand{\Iab}{\mathcal{J}}
\newcommand{\Trb}{\overline{T}^\text{rot}}

\newcommand{\llangle}{\langle\!\langle}
\newcommand{\rrangle}{\rangle\!\rangle}

\begin{document}
\title{Sonine approximation for  collisional moments of  granular gases of inelastic rough spheres}
\author{Andr\'es Santos}
\email{andres@unex.es}
\homepage{http://www.unex.es/eweb/fisteor/andres/}
\affiliation{Departamento de F\'{\i}sica, Universidad de Extremadura, E-06071
Badajoz, Spain}
\author{Gilberto M.  Kremer}
\email{kremer@fisica.ufpr.br}
\author{Marcelo dos Santos}
\affiliation{Departamento de F\'{\i}sica, Universidade Federal do Paran\'a, 81531-990 Curitiba, Brazil}

\begin{abstract}
We consider a dilute granular gas of hard spheres colliding inelastically with coefficients of normal and tangential restitution $\alpha$ and $\beta$, respectively.
The basic quantities characterizing the distribution function $f(\mathbf{v},\bm{\omega})$ of linear ($\mathbf{v}$) and angular ($\bm{\omega}$) velocities are the second-degree moments defining the  translational ($T^\text{tr}$) and rotational ($T^\text{rot}$) temperatures. The deviation of $f$ from the Maxwellian distribution parameterized by $T^\text{tr}$ and $T^\text{rot}$ can be measured by the cumulants associated with the fourth-degree velocity moments. The main objective of this paper is the evaluation of the collisional rates of change  of these second- and fourth-degree moments by means of a Sonine approximation. The results are subsequently applied to the computation of the temperature ratio $T^\text{rot}/T^\text{tr}$ and the cumulants of two paradigmatic states: the homogeneous cooling state and the homogeneous steady state driven by a white-noise stochastic thermostat. It is found in both cases that the Maxwellian approximation for the temperature ratio does not deviate much from  the Sonine prediction. On the other hand, non-Maxwellian properties measured by the cumulants cannot be ignored,  especially  in the homogeneous cooling state for medium and small roughness. In that state, moreover, the cumulant directly related to the translational velocity differs in the quasi-smooth limit $\beta\to -1$   from that of pure smooth spheres ($\beta=-1$). This singular behavior is directly related to the unsteady character of the homogeneous cooling state and thus it is absent in the stochastic thermostat case.
\end{abstract}


\date{\today}
\maketitle

\section{Introduction\label{sec1}}

Among the many topics in the kinetic theory of gases uncovered by Carlo Cercignani during his long and fruitful scientific career it is mandatory to mention the kinetic theory of \emph{inelastic} particles, a field he substantially contributed to during the last decade of his life.\cite{CCG00,C01,CIS01,C02,BC02,BC03,BCG03,BCG08,BCG09} With this paper we wish to pay a modest tribute to Carlo Cercignani's accomplishments in this field.

The most frequently used physical model of a granular fluid consists of a system of many inelastic and \emph{smooth} hard spheres with a constant coefficient of normal restitution $\alpha$.\cite{G03} On the other hand, the macroscopic nature of the grains makes the influence of \emph{friction} when two particles collide practically unavoidable.\cite{JR85,LS87,C89,L91,LN94,GS95,GNB05,L96,ZVPSH98,HZ98,ML98,LHMZ98,HHZ00,AMZ01,MHN02,CLH02,JZ02,PZMZ02,L95,MSS04,Z06,BPKZ07,KBPZ09,SKG10,S10a,S10b} {}From a more fundamental point of view, the existence of collisional friction is important to unveil the inherent breakdown of energy equipartition in granular fluids, even in homogeneous and isotropic states.

The simplest model accounting for friction during collisions assumes, apart from a constant coefficient of normal restitution $\al$, a {constant} coefficient of tangential restitution $\beta$.\cite{JR85,LS87} While $\alpha$ is a positive quantity smaller than or equal to 1 (the value $\alpha=1$ corresponding to elastic spheres), the parameter $\beta$ lies in the range between $-1$ (perfectly smooth spheres) to $1$  (perfectly rough spheres). The total kinetic energy is not conserved in a collision, unless  $\alpha=1$ \emph{and} $ \beta=\pm 1$.  As a consequence, many of the papers in the literature assume that the spheres are nearly smooth and nearly elastic.\cite{C89,L91,LN94,L96,ZVPSH98,JZ02,PZMZ02,GNB05}

\begin{figure}
  \includegraphics[width=.5\columnwidth]{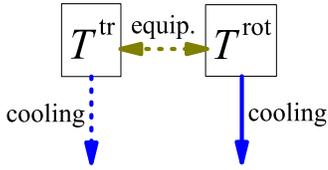}
\caption{(Color online) Scheme on the two classes of contributions (equipartition rates and cooling rates) to the energy production rates $\zt$ and $\zr$ characterizing the effect of collisions on $\Tt$ and $\Tr$, respectively. The terms represented by dotted arrows are absent in the case of perfectly smooth spheres ($\be=-1$).} \label{fig1}
\end{figure}

The theoretical study of a granular gas is usually undertaken by employing tools already developed in nonequilibrium statistical mechanics and kinetic theory of normal gases. In particular, one can introduce the one-body distribution function $f(\rr,\vv,\ww;t)$, where $\vv$ and $\ww$ are the velocity of the center of mass and the angular velocity, respectively, of a particle. {}From the second-degree velocity moments of the distribution function it is straightforward to define (granular) translational and rotational  temperatures, $\Tt$ and $\Tr$ (see Sec.\ \ref{sec2}). The rates of change of these two quantities produced by collisions define the energy production rates $\zt$ and $\zr$ as
\beq
\zt=-\frac{1}{\Tt}\left(\frac{\partial\Tt}{\partial t}\right)_{\text{coll}},\quad \zr=-\frac{1}{\Tr}\left(\frac{\partial\Tr}{\partial t}\right)_{\text{coll}}.
\label{1.1}
\eeq
The collisional energy production  rates $\zt$ and $\zr$  do not have a definite sign. They can be decomposed into two classes of terms:\cite{S10a} \emph{equipartition} rates and \emph{cooling} rates (see Fig.\ \ref{fig1}). The equipartition terms, which exist even when energy is conserved by collisions ($\al=1$ and $\be=\pm 1$), tend to make temperatures equal.\cite{S10a,UKAZ09}. Therefore, they can be positive or negative depending essentially on the sign of the temperature difference $\Tt-\Tr$. On the other hand, the genuine cooling terms reflect the collisional energy dissipation and thus they are positive if $\al<1$ and/or $|\be|<1$, vanishing otherwise. Only the cooling terms in $\zt$ and $\zr$ contribute to the net cooling rate  $\zeta=(\zt\Tt+\zr\Tr)/(\Tt+\Tr)$. Both $\zt$ and $\zr$ are functionals of $f$ and therefore they formally depend on all the moments of $f$, not just on $\Tt$ and $\Tr$.

In an extensive paper,\cite{GS95} Goldshtein and Shapiro undertook the task of evaluating the collisional energy production rates $\zt$ and $\zr$ by using a two-temperature \emph{Maxwellian} approximation for the distribution function, namely
\beqa
f(\vv,\ww)&\to & f_M(\vv,\ww)=n\left(\frac{m I}{4\pi^2\Tt\Tr}\right)^{3/2}\nn
&&\times\exp\left[-\frac{m(\vv-\mathbf{u})^2}{2\Tt}-\frac{I\omega^2}{2\Tr}\right].
\label{1.11}
\eeqa
Here, $n$ and $\mathbf{u}$ are  the number density and the flow velocity, respectively, of the gas and $m$ and $I$ are the mass and the moment of inertia, respectively, of a particle. The mean angular velocity has been assumed to vanish.\cite{GS95}
The final results for the energy production rates in the Maxwellian approximation are \cite{Z06,SKG10}
\beqa
\zt&=&\frac{5}{12}\left[1-\al^2+\frac{\q}{1+\q}\left(1-\be^2\right)\right.\nn
&&\left.+\frac{\q}{(1+\q)^2}\left(1+\be\right)^2\left(1-\theta\right)\right]\nu,
\label{1.3}
\eeqa
\beq
\zr=\frac{5}{12}\frac{1+\be}{1+\q}\left[1-\be-\frac{\q}{1+\q}\left(1+\be\right)\frac{1-\theta}{\theta}\right]\nu,
\label{1.4}
\eeq
where
\beq
\theta\equiv\frac{\Tr}{\Tt}
\label{1.6}
\eeq
is the rotational/translational temperature ratio,
\beq
\q\equiv \frac{4I}{m\sigma^2}
\label{1.7}
\eeq
is the dimensionless  moment of inertia ($\sigma$ being the  diameter of a particle), and
\beq
\nu\equiv\frac{16}{5}\sigma^2 n\sqrt{\pi \Tt/m}
\label{1.5}
\eeq
is an effective collision frequency. The expressions within the Maxwellian approximation but with a non-zero mean angular velocity can be found in Ref.\ \onlinecite{S10b}. Furthermore, the more general expressions for mixtures were derived in Ref.\ \onlinecite{SKG10}.

Equations \eqref{1.3} and \eqref{1.4} have been applied to the so-called homogeneous cooling state (HCS).\cite{GS95,HHZ00} {}From the condition $\lim_{t\to\infty}\theta(t)=\text{const}$ one gets  $\zt=\zr$, yielding a quadratic equation for the asymptotic temperature ratio $\theta$ whose physical solution is
\beq
\theta=\sqrt{1+C^2}+C
\label{1.8}
\eeq
with
\beq
C\equiv \frac{1+\kappa}{2\kappa(1+\beta)}\left[(1+\kappa)\frac{1-\alpha^2}{1+\beta}-(1-\kappa)(1-\beta)\right].
\label{1.9}
\eeq
The time evolution of the ratio $\Tr/\Tt$ toward the HCS asymptotic value \eqref{1.8} has been widely analyzed, both theoretically and by means of molecular dynamics, by Luding, Zippelius, and co-workers.\cite{HZ98,ML98,LHMZ98,HHZ00,AMZ01,CLH02,Z06}

An even simpler application of Eqs.\ \eqref{1.3} and \eqref{1.4} corresponds to the case of a homogeneous and isotropic granular gas kept in a nonequilibrium steady state by a white-noise thermostat.\cite{WM96,W96,SBCM98,vNE98,MS00} We will refer to this situation as the white-noise state (WNS). The steady-state condition $\zr=0$ simply yields
\beq
\theta=\q\frac{1+\beta}{1-\beta+2\q}.
\label{1.10}
\eeq

Despite the crudeness of the Maxwellian approximation given by Eq.\ \eqref{1.11}, Eq.\ \eqref{1.9} does a very good job when compared with computer simulations for the HCS.\cite{LHMZ98,Z06} The same is expected to hold for Eq.\ \eqref{1.10} in the WNS case. On the other hand, the production rates $\zt$ and $\zr$, being nonlinear functionals of $f$,  can be expected to be influenced by non-Maxwellian features of $f$, thus deviating (even if only slightly) from Eqs.\ \eqref{1.3} and \eqref{1.4}. The basic non-Maxwellian features of a velocity distribution function $f(\vv,\ww)$ are the existence of non-zero cumulants. The most physically interesting  cumulants are $a_{20}$, $a_{11}$, and $a_{02}$, defined as
\beq
\langle (\vv-\mathbf{u})^4\rangle=\frac{15}{4}\left(\frac{2\Tt}{m}\right)^2(1+a_{20}),
\label{1.12}
\eeq
\beq
\langle (\vv-\mathbf{u})^2\omega^2\rangle=\frac{9}{4}\left(\frac{2\Tt}{m}\right)\left(\frac{2\Tr}{I}\right)(1+a_{11}),
\label{1.13}
\eeq
\beq
\langle \omega^4\rangle=\frac{15}{4}\left(\frac{2\Tr}{I}\right)^2(1+a_{02}).
\label{1.14}
\eeq
Here the angular brackets denote average values defined as
\beq
\langle\psi(\vv,\ww)\rangle\equiv\frac{1}{n}\int d\vv\int d\ww\, \psi(\vv,\ww)f(\vv,\ww).
\label{1.15}
\eeq

The objectives of this paper are: (a) to evaluate the second-degree collisional moments $\zt$ and $\zr$ in a Sonine approximation that includes the cumulants $a_{20}$, $a_{11}$, and $a_{02}$; (b) to evaluate the three fourth-degree collisional moments related to the moments defined by Eqs.\ \eqref{1.12}--\eqref{1.14} in the same Sonine approximation; and (c) to apply the results to both the HCS and the WNS in order to ``refine''  Eqs.\ \eqref{1.8} and \eqref{1.10}, and estimate $a_{20}$, $a_{11}$, and $a_{02}$ in those states. The method will be similar to that already worked out in the case of smooth spheres.\cite{vNE98,MS00,SM09}

This paper is organized as follows.
The collision rules and the Boltzmann equation for a gas of inelastic and rough hard spheres are presented in Sec.\ \ref{sec2}.
The Sonine approximation is constructed in Sec.\ \ref{sec3}, where the derived expressions for the collisional moments are written down.
Sections \ref{sec4} and \ref{sec5} deal with the application of the results to the HCS and the WNS, respectively. The paper ends with a brief discussion in Sec.\ \ref{sec6}.

\section{Collision rules and Boltzmann equation\label{sec2}}
\subsection{Collision rules}
Let us consider a granular gas made of inelastic rough hard spheres of mass $m$, diameter $\sigma$, and moment of inertia $I$. In this section we first derive the  rules for a binary collision between two spheres with precollisional center-of-mass velocities $(\vv_1,\vv_2)$ and angular velocities ($\ww_1,\ww_2)$.

Let us denote by $\gh=\cca-\ccb$ the precollisional relative velocity of the center of mass of both spheres and by  $\kk\equiv (\mathbf{r}_2-\mathbf{r}_1)/|\mathbf{r}_2-\mathbf{r}_1|$  the unit vector pointing from the center of sphere $1$ to the center of sphere $2$. The precollisional velocities of  the points of the spheres which
are in contact during the collision are
\beq
\wwwa=\cca-\frac{\da}{2}\kk\x\wwa,\quad \wwwb=\ccb+\frac{\da}{2}\kk\x\wwb,
\label{w}
\eeq
the corresponding relative velocity being
\beq
\wwwab=\gh-\kk\x\SSab,\quad \SSab\equiv \frac{\da}{2}(\wwa+\wwb).
\label{Sab}
\eeq

Conservation of linear and angular momenta yields\cite{Z06}
\beq
\cca'+\ccb'=\cca+\ccb,
\label{momentum}
\eeq
\begin{subequations}
\label{angular}
\beq
I\wwa'-m \frac{\da}{2}\kk\x\cca'=I\wwa-m \frac{\da}{2}\kk\x\cca,
\eeq
\beq
I\wwb'+m \frac{\da}{2}\kk\x\ccb'=I\wwb+m \frac{\da}{2}\kk\x\ccb,
\eeq
\end{subequations}
where the primes denote postcollisional values.
Equations \eqref{momentum} and \eqref{angular} imply that
\beq
\cca'=\cca-\JJ,\quad \ccb'=\ccb+\JJ,
\label{13b}
\eeq
\beq
\wwa'=\wwa-\frac{m\da}{2I}\kk\x\JJ, \quad \wwb'=\wwb-\frac{m\da}{2I}\kk\x\JJ,
\label{14w}
\eeq
where $m\JJ$ is the impulse exerted by particle $1$ on particle $2$.
Therefore,
\beq
\gh'=\gh-2\JJ,\quad \wwwab'=\wwwab-2\JJ+\frac{2}{\q}\kk\x\left(\kk\x\JJ\right),
\label{wwwab'}
\eeq
where the dimensionless moment of inertia $\q$ is defined by  Eq.\ \eqref{1.7}. It varies from zero to a maximum value of $\frac{2}{3}$, the former corresponding to a concentration of the mass at the center of the sphere, while the latter value corresponds to a concentration of the mass on the surface of the sphere. The value $\q=\frac{2}{5}$ refers to a uniform mass distribution.

To close the collision rules, we need to express $\JJ$ in terms of the precollisional velocities and the unit vector $\kk$. To that end, let us relate the normal (i.e., parallel to $\kk$) and tangential (i.e., orthogonal to $\kk$) components of the relative velocities $\wwwab$ and $\wwwab'$ by
\beq
\kk\cdot \wwwab'=-\al \kk\cdot \wwwab,\quad \kk\x \wwwab'=-\be \kk\x \wwwab.
\label{restitution}
\eeq
Here, as said in Sec.\ \ref{sec1}, $\al$ and $\be$ are the coefficients of normal and tangential restitution, respectively.
The former coefficient ranges from $\al=0$ (perfectly inelastic particles) to $\al=1$ (perfectly elastic particles), while the latter runs from $\be=-1$ (perfectly smooth particles) to $\be=1$ (perfectly rough particles).
{A more realistic model consists of assuming that $\beta$ is a function of the angle between $\wwwab$ and $\kk$,\cite{HHZ00} thus accounting for Coulomb friction. In this paper, however, we will assume a constant $\beta$.}

Inserting the second equality of Eq.\ \eqref{wwwab'} into Eq.\ \eqref{restitution} one  gets
\beq
\kk\cdot\JJ=\at \kk\cdot\wwwab,\quad \kk\x\JJ=\bt \kk\x\wwwab,
\eeq
where the following abbreviations were introduced:
\beq
\at\equiv\frac{1+\al}{2},\quad\bt\equiv\frac{\q}{1+\q}\frac{1+\be}{2}.
\label{20a}
\eeq
Therefore,
\beq
\JJ=\at (\gh\cdot\kk)\kk+\bt\left[\gh-(\gh\cdot\kk)\kk-\kk\x\SSab\right].
\label{15}
\eeq

Equations \eqref{13b},  \eqref{14w}, and \eqref{15} express the postcollisional velocities $(\vv_1',\ww_1',\vv_2',\ww_2')$ in terms of the precollisional velocities $(\vv_1,\ww_1,\vv_2,\ww_2)$ and the unit vector $\kk$. In the special case of perfectly smooth spheres ($\be=-1$ or, equivalently, $\bt=0$) one has $\kk\x\JJ=\mathbf{0}$, so that $\ww_1'=\ww_1$ and $\ww_2'=\ww_2$.

The collisional change of the total (translational plus rotational) kinetic energy is
\beqa
E_{12}'-E_{12}&=&-\frac{m}{4}\left(1-\al^2\right)(\kk\cdot\gh)^2
\nn
&&-\frac{m}{4}\frac{\q}{1+\q}\left(1-\be^2\right)\left[\gh\right.\nn
&&\left.-\kk\x\SSab-(\gh\cdot\kk)\kk\right]^2,
\label{29E}
\eeqa
where
\beq
E_{12}\equiv\frac{m }{2}\ca^2+\frac{m }{2}\cb^2+\frac{I}{2}\wa^2+\frac{I}{2}\wb^2.
\label{Z2}
\eeq
The right-hand side of Eq.\ \eqref{29E} is a negative definite quantity. Thus,  energy is conserved only if the particles are elastic  ($\al=1$)  and  either perfectly smooth ($\be=-1$) or perfectly rough ($\be=1$). Otherwise, $E_{12}'<E_{12}$ and kinetic energy is dissipated upon collisions.

Equations \eqref{13b}, \eqref{14w}, and \eqref{15} give the \emph{direct} collisional rules.
For a restituting encounter the pre- and postcollisional
velocities are denoted by $(\cca'',\wwa'',\ccb'',\wwb'')$ and $(\cca,\wwa,\ccb,\wwb)$, respectively, and the collision vector
by $\kk''=-\kk$. It is easy to verify that the relationship $\gh\cdot\kk''=-\al \gh''\cdot\kk''=-\gh\cdot\kk$ holds. Analogously,
$\kk''\x\wwwab=-\be \kk''\x\wwwab''=-\kk\x\wwwab$. As a consequence, the \emph{restituting} collision rules are
\beq
\cca''=\cca-\JJw,\quad \ccb''=\ccb+\JJw,
\label{13brest}
\eeq
\beq
\wwa''=\wwa-\frac{m\da}{2I}\kk\x\JJw, \quad \wwb''=\wwb-\frac{m\da}{2I}\kk\x\JJw,
\label{14rest}
\eeq
where
\beq
\JJw=\frac{\at}{\al} (\gh\cdot\kk)\kk+\frac{\bt}{\be}\left[\gh-(\gh\cdot\kk)\kk-\kk\x\SSab\right].
\label{15rest}
\eeq

 The modulus of the Jacobian of the transformation between pre- and postcollisional velocities is
\beq
\left|\frac{\partial(\cca',\wwa',\ccb',\wwb')}{\partial(\cca,\wwa,\ccb,\wwb)}\right|=
\left|\frac{\partial(\cca,\wwa,\ccb,\wwb)}{\partial(\cca'',\wwa'',\ccb'',\wwb'')}\right|={\al\be^2}.
\label{Jacob}
\eeq
Furthermore, the relationship between volume elements in velocity space reads
\beq
|\gh''\cdot\kk''|d\cca''d\wwa''d\ccb''d\wwb''=\frac{|\gh\cdot\kk|}{\al^2\be^2}d\cca d\wwa d\ccb d\wwb.
\label{1.16}
\eeq

\subsection{Boltzmann equation}
If the granular gas is dilute enough the velocity distribution function $f(\rr,\cc,\ww;t)$ obeys the Boltzmann equation\cite{GS95,BDS97}
\beq
\partial_t f+\cc\cdot \nabla f={J[\cc,\ww|f]},
\label{2.1}
\eeq
where the collision operator is
\beqa
{J\left[{\bf v}_{1},\ww_1|f\right]}&=&\sigma^{2} \int
d{\bf v}_{2}\int d\ww_2\int_+ d\widehat{\bm{\sigma}} \,\left(\mathbf{\gh}\cdot\widehat{\bm{\sigma}}\right)\nn
&&
\times\left(
\frac{1}{\alpha^2\beta^2}f_1''f_2''-f_{1}f_2\right).
\label{2.2}
\eeqa
Here the subscript $+$ in the integral over $\kk$ means the constraint $\mathbf{\gh}\cdot\widehat{\bm{\sigma}}>0$ and we have employed the short-hand notation $f_1''\equiv f(\vv_1'',\ww_1'')$ and so on.

Given an arbitrary function $\psi(\vv,\ww)$, its average value is defined by Eq.\ \eqref{1.15}. The associated collisional rate of change is ${\langle \psi\rangle^{-1} \Iab[\psi|f]}$, where the collisional quantity ${\Iab[\psi(\cc,\ww)|f]}$ is defined by
\beqa
{\Iab[\psi|f]}&\equiv&\int d\cca\int d\wwa\, \psi(\cca,\wwa) J[\cca,\wwa|f]\nn
&=&\frac{\da^2}{2}\int d\cca\int d\wwa \int d\ccb \int d\wwb\int_+ d\kk\,
\nn&&\x(\gh\cdot\kk)
f_1f_2\left(\psi_1'+\psi_2'-\psi_1-\psi_2\right),\nn
\label{3}
\eeqa
where in the last step we have carried out a standard change of variables.

Here we are especially concerned with the \emph{partial} temperatures associated with the translational and rotational degrees of freedom:
\beq
\Tt=\frac{m}{3}\langle (\cc-\mathbf{u})^2\rangle,\quad \Tr=\frac{I}{3}\langle \omega^2\rangle,
\label{III.11}
\eeq
where $\mathbf{u}\equiv \langle\cc\rangle$ is the flow velocity.
The corresponding energy production rates are defined as
\beq
\zt\equiv -\frac{m}{3n\Tt}{\Iab[(\cc-\mathbf{u})^2|f]},\quad \zr\equiv -\frac{I}{3n\Tr}{\Iab[\omega^2|f]}.
\label{54}
\eeq
The \emph{total} temperature and its corresponding \emph{cooling} rate are
\beq
T=\frac{\Tt+\Tr}{2},
\eeq
\beq
\zeta\equiv \frac{\zt\Tt+\zr\Tr}{\Tt+\Tr}.
\label{3.1}
\eeq

It is worthwhile remarking that, instead of $\Tr$, we could have alternatively adopted
$\Trb=({I}/{3})\langle\left(\bm{\omega}-\langle\bm{\omega}\rangle\right)^2\rangle=\Tr\left(1-X\right)$, with
$ X\equiv {\q m\sigma^2\langle\bm{\omega}\rangle^2}/{12\Tr}$,
as the definition of the rotational temperature.
However, a disadvantage of this alternative choice  is that, in contrast to the cooling rate $\zeta$ defined by Eq.\ \eqref{3.1},  the alternative ``cooling'' rate  $\overline{\zeta}$ associated with the alternative total temperature   $\overline{T}=\frac{1}{2}(\Tt+\Trb)=T-\frac{1}{2}\Tr X$ is not positive definite and in fact becomes negative in the perfectly elastic and rough case ($\alpha=1$, $\beta=1$).\cite{S10b}

Making use of the collision rules given by Eqs.\ \eqref{13b}, \eqref{14w}, and \eqref{15}, and after performing the integration over $\kk$, one gets\cite{SKG10}
\beqa
\zt&=&\frac{5\sqrt{\pi}\nu}{96(\Tt/m)^{3/2}}\Bigg[\left(\at(1-\at)+\bt(1-\bt)\right){\llangle v_{12}^3\rrangle}\nn
&&-\frac{\bt^2}{2}{\llangle 3v_{12} S_{12}^2- v_{12}^{-1} \left(\gh\cdot\SSab\right)^2\rrangle}\Bigg],
\label{3.2}
\eeqa
\beqa
\zr&=&\frac{5\sqrt{\pi}\nu}{96(\Tt/m)^{3/2}}\frac{\bt}{\theta}\Bigg[
\frac{1}{2}\left(1-\frac{\bt}{\q}\right){\llangle 3 v_{12} S_{12}^2}\nn
&&{- v_{12}^{-1} \left(\gh\cdot\SSab\right)^2\rrangle}-\frac{\bt}{\q}{\llangle v_{12}^3\rrangle}\Bigg],
\label{3.3}
\eeqa
\beqa
\zeta&=&\frac{5\sqrt{\pi}\nu}{384(\Tt/m)^{3/2}}\frac{1}{1+\theta}\Bigg[(1-\al^2){\llangle v_{12}^3\rrangle}\nn
&&+\frac{\q}{1+\q}\frac{1-\be^2}{2}\Big(2{\llangle v_{12}^3\rrangle+\llangle 3v_{12} S_{12}^2}\nn
&&{- v_{12}^{-1} \left(\gh\cdot\SSab\right)^2\rrangle}\Big)\Bigg].
\label{3.4}
\eeqa
In these equations,  $\theta$ is the temperature ratio defined by Eq.\ \eqref{1.6}, $\nu$ is the collision frequency defined by Eq.\ \eqref{1.5},
and
\beqa
{\llangle \psi(\gh,\SSab)\rrangle}&\equiv &\frac{1}{n^2}\int d\cca\int d\wwa \int d\ccb\int d\wwb\nn
&&\times \psi(\gh,\SSab)
f(\cca,\wwa)f(\ccb,\wwb)\nn
\label{3.5}
\eeqa
are two-body averages. Use has been made of Eq.\ \eqref{20a} upon obtaining  Eq.\ \eqref{3.4} from Eqs.\ \eqref{3.2} and \eqref{3.3}.
It is worthwhile emphasizing that Eqs.\ \eqref{3.2}--\eqref{3.4} are \emph{exact} in the framework of the Boltzmann equation.

\section{Sonine approximation for second- and fourth-degree collisional moments\label{sec3}}
Equations \eqref{3.2} and \eqref{3.3} express the translational and rotational energy production rates as functionals of $f$ through two independent two-body averages of the {form} given by Eq.\ \eqref{3.5}. If $f$ is replaced by the Maxwellian approximation Eq.\ \eqref{1.11} one gets Eqs.\ \eqref{1.3} and \eqref{1.4}. As said in Sec.\ \ref{sec1} we want to go beyond such a Maxwellian approximation.

To proceed, it is convenient to  introduce the  dimensionless velocities
\beq
\mathbf{c}\equiv\frac{\mathbf{v}-\mathbf{u}}{\sqrt{2\Tt/m}},
\quad \mathbf{w}\equiv\frac{\bm{\omega}}{\sqrt{2\Tr/I}},
\label{7a}
\eeq
and the dimensionless distribution function
\beq
\phi(\mathbf{c},\mathbf{w})\equiv \frac{1}{n}\left(\frac{4\Tt\Tr}{m I}\right)^{3/2}f(\mathbf{v},\bm{\omega}).
\label{7b}
\eeq
In terms of the reduced translational and rotational velocities, the collision rules given by Eqs.\ \eqref{13b}, \eqref{14w}, and \eqref{15} become
\beq
\mathbf{c}_{1}'=\mathbf{c}_{1}-\JJ^*, \quad \mathbf{c}_{2}'=\mathbf{c}_{2}+\JJ^*,
\label{25}
\eeq
\beq
\mathbf{w}_1'=\mathbf{w}_1-\frac{1}{\sqrt{\kappa\theta}}\widehat{\bm{\sigma}}\times\JJ^*,
\quad \mathbf{w}_2'=\mathbf{w}_2-\frac{1}{\sqrt{\kappa\theta}}\widehat{\bm{\sigma}}\times\JJ^*,
\label{26}
\eeq
\beqa
\JJ^*&=&\at\left(\mathbf{c}_{12}\cdot\widehat{\bm{\sigma}}\right)\widehat{\bm{\sigma}}+
\bt\Bigg[\mathbf{c}_{12}-\left(\mathbf{c}_{12}\cdot\widehat{\bm{\sigma}}\right)\widehat{\bm{\sigma}}\nn
&&
-\sqrt{\frac{\theta}{\kappa}}\widehat{\bm{\sigma}}\times\left(\mathbf{w}_1+\mathbf{w}_2\right)\Bigg].
\label{27}
\eeqa

Let us now specialize to \emph{isotropic} states. The latter condition implies that {the scalar function $\phi(\mathbf{c},\mathbf{w})$ is  invariant under orthogonal transformations, including those with determinant equal to $+1$ (rotations) or $-1$ (reflections). This means that $\phi(\mathbf{c},\mathbf{w})$ is
actually a function of the three scalar quantities $c^2=\mathbf{c}\cdot\mathbf{c}$, $w^2=\mathbf{w}\cdot\mathbf{w}$, and $(\mathbf{c}\cdot \mathbf{w})^2$.} We do not need to assume that the state is either homogeneous or stationary.
Here we focus on the following second- and fourth-degree moments: {$\langle c^2\rangle$, $\langle w^2\rangle$, $\langle c^4\rangle$, $\langle c^2w^2\rangle$, and $\langle w^4\rangle$.}
By construction,
\beq
\langle c^2\rangle=\langle w^2\rangle=\frac{3}{2}.
\eeq
In the Maxwellian approximation \eqref{1.11}, i.e.,
\beq
\phi(\mathbf{c},\mathbf{w})\to\phi_M(\mathbf{c},\mathbf{w})=\pi^{-3}e^{-c^2-w^2},
\label{16}
\eeq
one has
\beq
{\langle c^4\rangle\to\frac{15}{4},\quad \langle c^2w^2\rangle\to\frac{9}{4},\quad \langle w^4\rangle\to\frac{15}{4}.}
\label{18}
\eeq
In general, however, $\phi\neq\phi_M$ and the above equalities are not verified. {This can be characterized by the \emph{cumulants}}
\beq
{a_{20}=\frac{4}{15}\langle c^4\rangle-1},
\label{19a}
\eeq
\beq
 a_{11}=\frac{4}{9}\langle c^2w^2\rangle-1,
\label{19b}
\eeq
\beq
 a_{02}=\frac{4}{15}\langle w^4\rangle-1.
\label{19c}
\eeq
Note that {Eqs.\ \eqref{19a}--\eqref{19c} are}  equivalent to Eqs.\ \eqref{1.12}--\eqref{1.14}.

Let us define the  {\emph{collisional} moments $\mu_{pq}$ (with $p,q=\text{even}$) as}
\beq
{\mu_{pq}=-\int d\mathbf{c}\int d\mathbf{w}\, c^pw^q{J^*[\mathbf{c},\mathbf{w}|\phi]},}
\label{10b}
\eeq
where  the dimensionless collision operator $J^*$ is defined similarly to Eq.\ \eqref{2.2}, except that one must formally take $\sigma=1$ and the collision rules are given by Eqs.\ \eqref{25}--\eqref{27}. The energy production rates $\zt$ and $\zr$ are directly related to the collisional moments $\mu_{20}$ and $\mu_{02}$ by
\beq
\label{3.6}
\zt=\frac{5\nu}{12\sqrt{2\pi}}\mu_{20},\quad \zr=\frac{5\nu}{12\sqrt{2\pi}}\mu_{02}.
\eeq
Analogously, the total cooling rate is
\beq
\zeta=\frac{5\nu}{12\sqrt{2\pi}}\frac{\mu_{20}+\mu_{02}}{1+\theta}.
\label{3.9bis}
\eeq

The primary objective in this section is to get \emph{estimates} of the second-degree collisional moments  {$\mu_{20}$ and $\mu_{02}$}, and of the fourth-degree collisional moments $\mu_{40}$, $\mu_{22}$, and $\mu_{04}$ in terms of the temperature ratio $\theta$ and the cumulants {$a_{20}$, $a_{11}$, and  $a_{02}$.} To that end, we first express the distribution function $\phi$ by the first few terms in its Sonine expansion,
\beqa
\phi(\mathbf{c},\mathbf{w})&\approx& \phi_M(\mathbf{c},\mathbf{w})\left[{1+a_{20}S_{\frac{1}{2}}^{(2)}(c^2)+a_{02}S_{\frac{1}{2}}^{(2)}(w^2)}\right.\nn
&&\left.+a_{11}S_{\frac{1}{2}}^{(1)}(c^2)S_{\frac{1}{2}}^{(1)}(w^2)\right].\nn
\label{20}
\eeqa
The Sonine polynomials in Eq.\ \eqref{20} are
\beq
S_{\frac{1}{2}}^{(1)}(x)=\frac{3}{2}-x,\quad S_{\frac{1}{2}}^{(2)}(x)=\frac{1}{8}\left(15-20x+4x^2\right).
\label{21}
\eeq
In principle, apart from the moments $\langle c^4\rangle$, $\langle c^2 w^2\rangle$, and $\langle w^4\rangle$, {the other  independent  fourth-degree moment $\langle (\mathbf{c}\cdot\mathbf{w})^2\rangle$} should be represented in the truncated  expansion \eqref{20}. However, for simplicity, {it is assumed here that}
\beq
{\langle (\mathbf{c}\cdot\mathbf{w})^2\rangle=\frac{1}{3}\langle c^2w^2\rangle=\frac{3}{4}(1+a_{11}).}
\label{25.2}
\eeq
{This implies that the study of the \emph{orientational} correlation between $\mathbf{c}$ and $\mathbf{w}$ is not addressed in this paper. {}From that point of view, our approach is complementary to that of Refs.\ \onlinecite{BPKZ07,KBPZ09}, where it was assumed that $a_{20}=a_{11}=a_{02}=0$ but  $\langle (\mathbf{c}\cdot\mathbf{w})^2\rangle/\langle c^2w^2\rangle\neq \frac{1}{3}$.}

The second step consists of inserting the approximation defined by Eq.\ \eqref{20} into {Eq.\ \eqref{10b}}, and neglecting terms nonlinear in {$a_{20}$, $a_{11}$, and $a_{02}$.} After some algebra one gets the following expressions for the second-degree collisional moments:
\beqa
{\mu_{20}}&=&{4\sqrt{2\pi}\left[\left(\at(1-\at)+\bt(1-\bt)\right)\left(1+\frac{3a_{20}}{16}\right)\right.}\nn
&&{\left.-\theta\frac{\bt^2}{\kappa}\left(1-\frac{a_{20}}{16}+\frac{a_{11}}{4}\right)\right]},
\label{22}
\eeqa
\beqa
\mu_{02}&=&4\sqrt{2\pi}\frac{\bt}{\kappa}\left[\left(1-\frac{\bt}{\kappa}\right)\left(1-\frac{a_{20}}{16}+\frac{a_{11}}{4}\right)\right.\nn
&&\left.-\frac{\bt}{\theta}\left(1+\frac{3a_{20}}{16}\right)\right].
\label{23}
\eeqa
Thus, Eq.\ \eqref{3.9bis} gives
\beqa
\zeta
&=&\frac{5\nu}{12(1+\theta)}\left[
\left(1-\al^2+\q\frac{1-\be^2}{1+\q}\right)\left(1+\frac{3a_{20}}{16}\right)\right.\nn
&&\left.+\theta\frac{1-\be^2}{1+\q}\left(1-\frac{a_{20}}{16}+\frac{a_{11}}{4}\right)\right].
\label{3.9}
\eeqa
Equations \eqref{22}--\eqref{3.9} can also be obtained from Eqs.\ \eqref{3.2}--\eqref{3.4} by taking into account that
\beq
{\llangle v_{12}^3\rrangle}\approx \frac{16}{\sqrt{2\pi}}\left(\frac{2\Tt}{m}\right)^{3/2}\left(1+\frac{3a_{20}}{16}\right),
\label{3.7}
\eeq
\beqa
{\llangle 3v_{12} S_{12}^2- v_{12}^{-1} \left(\gh\cdot\SSab\right)^2\rrangle}&\approx &
\frac{32\theta}{\q\sqrt{2\pi}}\left(\frac{2\Tt}{m}\right)^{3/2}\nn
&&\times\left(1-\frac{a_{20}}{16}+\frac{a_{11}}{4}\right)\nn
\label{3.8}
\eeqa
in the Sonine approximation \eqref{20}. This explains why the cumulant $a_{02}$, being related to $\langle \omega^4\rangle$, does not intervene in Eqs.\ \eqref{22}--\eqref{3.9}.

Of course, Eqs.\ \eqref{22} and \eqref{23} reduce to Eqs.\ \eqref{1.3} and \eqref{1.4}, respectively, by setting $a_{20}=a_{11}=0$. Moreover, Eq.\ \eqref{22} is consistent with van Noije and Ernst's derivation\cite{vNE98} for the smooth case ($\be=-1$).

The evaluation of the fourth-degree collisional moments $\mu_{40}$, $\mu_{22}$, and $\mu_{04}$ is much more involved. After carefully performing the calculations following several independent routes to check the results, we have found
\begin{widetext}
\beqa
\mu_{40}&=&16\sqrt{2\pi}
\left\{\at^3(2-\at)+\bt^3(2-\bt)-\at\bt (1-\at-\bt+\at\bt)+\frac{11}{8}(\at+\bt)-\frac{19}{8}(\at^2+\bt^2)\right.\nn
&&-\left[\at\bt \left(\frac{23}{15}-\at-\bt+\at\bt\right)-\frac{269}{120}(\at+\bt)+\frac{357}{120}(\at^2+\bt^2)-\at^3(2-\at)-\bt^3(2-\bt)\right]\frac{15a_{20}}{16}\nn
&&-\frac{11 \bt^2\theta}{8\kappa}\left(1+\frac{41a_{20}}{176 }+\frac{3a_{11}}{4}\right)+\frac{\bt^2\theta}{\kappa}\left[\at(1-\at)+2\bt(1-\bt)\right]\left(1+\frac{3a_{20}}{16}+\frac{3a_{11}}{4}\right)\nn
&&\left.-\frac{\bt^4 \theta^2}{\kappa^2}\left(1-\frac{a_{20}}{16}+\frac{a_{11}}{2}+\frac{a_{02}}{2}\right)\right\},
\label{28}
\eeqa
\beqa
\mu_{22}&=&
3\sqrt{2\pi }\left\{2\left[\at(1-\at)+\bt(1-\bt)-\frac{4\at\bt}{3\kappa}(1-\at)\left(1-\frac{\bt}{\kappa}\right)-\frac{8\bt^2}{3\kappa}\left(\frac{3}{4}-\bt-\frac{\bt}{\kappa}+2\frac{\bt^2}{\kappa}\right)\right]
\right.\nn
&&\times \left(1+\frac{3a_{20}}{16}+\frac{3a_{11}}{4}\right)+\frac{7\bt}{3\kappa}\left(1-\frac{\bt}{\kappa}\right)\left(1+\frac{29a_{20}}{112}\right)-\frac{\bt^2}{2\kappa\theta}a_{20}
-\frac{8\bt^2}{3\kappa\theta}\Bigg[\frac{9}{8}-\at(1-\at)\nn
&&-2\bt(1-\bt)
\Bigg]\left(1+\frac{15a_{20}}{16}\right)-\frac{\bt^2\theta}{3\kappa}\left[5-8 \frac{\bt}{\kappa}\left(1-\frac{\bt}{\kappa}\right)\right]a_{02}-8\frac{\bt^2\theta}{3\kappa}\left[1-2 \frac{\bt}{\kappa}\left(1-\frac{\bt}{\kappa}\right)\right]\nn
&&\left.\times \left(1-\frac{a_{20}}{16}+\frac{a_{11}}{2}\right)+\left[\frac{\bt}{\kappa}\left(\frac{37}{12}-2\bt-\frac{7\bt}{4\kappa}\right)+\at+\bt-\frac{4 \at\bt}{3\kappa}\right]a_{11}\right\},
\label{29}
\eeqa
\beqa
\mu_{04}&=&4\sqrt{2\pi }\frac{\bt}{\kappa}\left\{\left(1-\frac{\bt}{\kappa}\right)\left[5-4\frac{\bt}{\kappa}\left(1-\frac{\bt}{\kappa}\right)\right]\left(1-\frac{a_{20}}{16}\right)- \frac{\bt}{\theta}\left[5-8\frac{\bt}{\kappa}\left(1-\frac{\bt}{\kappa}\right)\right]\left(1+\frac{3a_{20}}{16}+\frac{3a_{11}}{4}\right)
\right.\nn
&&\left.-\frac{5}{2}\left(1-\frac{4 \bt}{5\kappa}\right)a_{11}-\frac{4\bt^3}{\kappa\theta^2}\left(1+\frac{15a_{20}}{16}\right)+
\left(5-\frac{13}{2}\frac{\bt}{\kappa}+4\frac{\bt^2}{\kappa^2} - 2\frac{\bt^3}{\kappa^3}\right)\left({a_{11}}+{a_{02}}\right)\right\}.
\label{30}
\eeqa
\end{widetext}

To the best of our knowledge, the collisional moments $\mu_{40}$, $\mu_{22}$, and $\mu_{04}$ have not been evaluated before, even in the Maxwellian approximation ($a_{20}=a_{11}=a_{02}=0$). The only exception is van Noije and Ernst's evaluation of $\mu_{40}$ in the smooth case,\cite{vNE98} to which Eq.\ \eqref{28} reduces by setting $\be=-1$. As an additional simple consistency test, we get $\mu_{22}=\frac{3}{2}\mu_{20}$ in the special case of smooth spheres ($\be=-1$) with $a_{11}=0$.

Equations {\eqref{22}, \eqref{23},} and \eqref{28}--\eqref{30} are the main results of this paper. In the next two sections they are applied to the HCS and the WNS.

\section{Application to the homogeneous cooling state\label{sec4}}
In the homogeneous free cooling state (HCS) the Boltzmann equation \eqref{2.1} becomes
\beq
\partial_t f(\vv,\ww;t)={J[\vv,\ww|f]}.
\label{4.3}
\eeq
As a consequence, the only mechanisms responsible for changes in the partial and total temperatures are collisions. More specifically,
\beq
\partial_t\Tt=-\zt \Tt,\quad \partial_t\Tr=-\zr \Tr,
\label{4.1}
\eeq
\beq
\partial_t T=-\zeta  T.
\label{4.2}
\eeq
The evolution equation for the temperature ratio $\theta=\Tr/\Tt$ is $\partial_t\theta =-(\zr-\zt)\theta$.

Carrying out the change to dimensionless  variables defined by Eqs.\ \eqref{7a} and \eqref{7b}, Eq.\ \eqref{4.3} can be rewritten as
\beq
\partial_s\phi+\frac{\mu_{20}}{3}\frac{\partial}{\partial\mathbf{c}}\cdot\left(\mathbf{c}\phi\right)
+\frac{\mu_{02}}{3}\frac{\partial}{\partial\mathbf{w}}\cdot\left(\mathbf{w}\phi\right)= {J^*[\mathbf{c},\mathbf{w}|\phi]},
\label{4.4}
\eeq
where $\partial_s\equiv (n\sigma^2\sqrt{{2\Tt}/{m}})^{-1}\partial_t$ and use has been made of Eq.\ \eqref{3.6}.
Taking moments in Eq.\ \eqref{4.4} we get
\beq
{-\partial_s \langle c^pw^q\rangle+\frac{1}{3}(p\mu_{20}+q\mu_{02})\langle c^pw^q\rangle=\mu_{pq}.}
\label{11}
\eeq

\begin{figure}
  \includegraphics[width=.95\columnwidth]{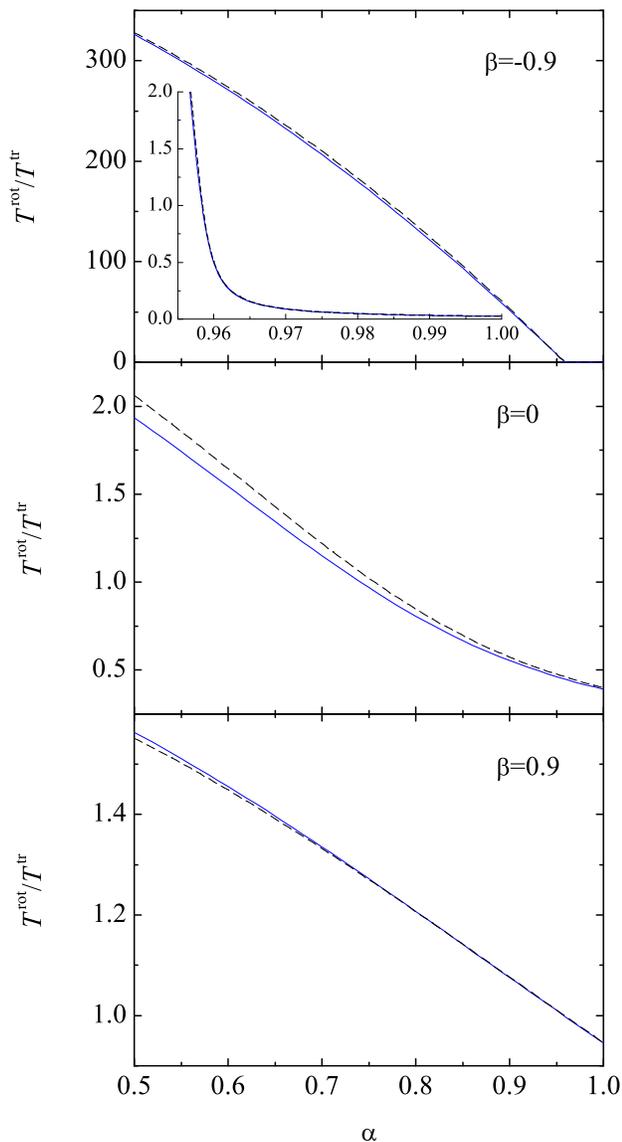}
\caption{(Color online) Plot of the HCS temperature ratio $\Tr/\Tt$ vs the coefficient of normal restitution $\alpha$ for $\beta=-0.9$ (top panel), $\beta=0$ (middle panel), and $\beta=0.9$ (bottom panel).
The inset in the top panel is a blow-up of the region $0.95\leq\al\leq 1$. The dashed and solid lines are the Maxwellian and Sonine approximations, respectively.} \label{fig2}
\end{figure}

\begin{figure}
  \includegraphics[width=.95\columnwidth]{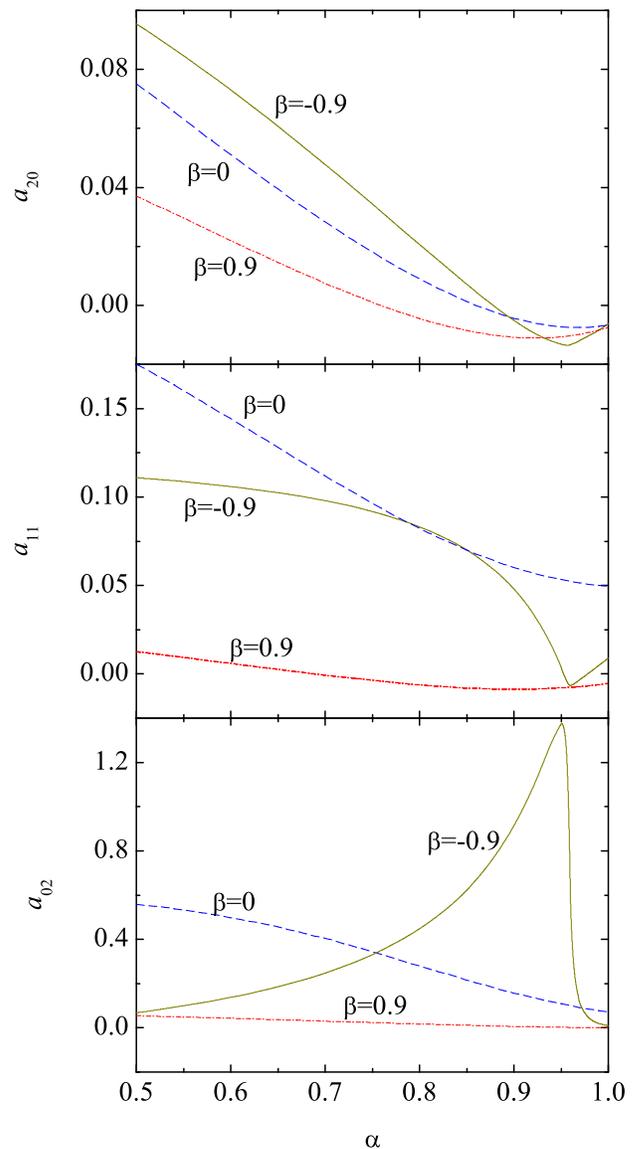}
\caption{(Color online) Plot of the HCS cumulants $a_{20}$ (top panel), $a_{11}$ (middle panel), and $a_{02}$ (bottom panel) vs the coefficient of normal restitution  $\al$ for $\beta=-0.9$ (solid lines), $\beta=0$ (dashed lines), and $\beta=0.9$ (dash-dotted lines).} \label{fig3}
\end{figure}

After a certain transient period, it is expected that the system reaches an asymptotic regime where all the time dependence of $f$ appears through one temperature (say $\Tt$) and the temperature ratio $\theta$ remains constant. This implies  a \emph{similarity} solution of Eq.\ \eqref{4.3} of the form given by Eq.\ \eqref{7b} with $\partial_s\phi=0$.
Moreover, the condition $\partial_t\theta=0$ implies $\zt=\zr=\zeta$ or, equivalently,
\beq
\mu_{20}=\mu_{02}.
\label{12}
\eeq
In the asymptotic regime, {Eq.\  \eqref{11} yields}
\beq
{5\mu_{20}=\frac{\mu_{40}}{1+a_{20}},}
\label{13}
\eeq
\beq
\frac{3}{2}(\mu_{20}+\mu_{02})=\frac{\mu_{22}}{1+a_{11}},
\label{14}
\eeq
\beq
5\mu_{02}=\frac{\mu_{04}}{1+a_{02}}.
\label{15bb}
\eeq

The objective now is to \emph{estimate} the temperature ratio $\theta$ and the cumulants {$a_{20}$, $a_{11}$, and $a_{02}$} in the HCS.
To that end, we insert the approximate expressions given by Eqs.\ {\eqref{22}, \eqref{23}, and} \eqref{28}--\eqref{30} into Eqs.\ \eqref{12}--\eqref{15bb},  neglecting again terms nonlinear in {$a_{20}$, $a_{11}$, and $a_{02}$.} Note that, for instance, Eq.\ \eqref{13} could also be written as $5\mu_{20}({1+a_{20}})={\mu_{40}}$.\cite{vNE98} However, the linearization process gives a result different from the one obtained from the form \eqref{13}, as discussed in Ref.\ \onlinecite{SM09}.  We have chosen the route \eqref{13} because it yields results more accurate for smooth spheres than the other one.\cite{MS00,SM09}

{{}From the linearized versions of Eqs.\ \eqref{12}--\eqref{14} one can obtain the three cumulants $a_{20}$, $a_{11}$, and $a_{02}$} as nonlinear functions of $\al$, $\be$, and $\theta$. Insertion into Eq.\ \eqref{15bb} yields an eighth-degree equation for $\theta$, whose physical solution is chosen as the one close to the solution \eqref{1.8} in the Maxwellian approximation. The final expressions are too cumbersome to be explicitly reproduced here but they are easy to deal with the help of a computer algebra system.

To illustrate the dependence of $\theta$,  $a_{20}$, $a_{11}$, and $a_{02}$ on both $\al$ and $\be$, we plot those quantities as functions of $\al$ for three representative values of the coefficient of tangential restitution: $\beta=-0.9$ (small roughness), $\beta=0$ (medium roughness), and $\beta=0.9$ (large roughness). In all the cases the density of the spheres has been assumed to be uniform, so that $\q=\frac{2}{5}$.
The results are displayed in Figs.\ \ref{fig2} and \ref{fig3}. In Fig.\ \ref{fig2} we observe that the Maxwellian approximation [cf. Eq.\ \eqref{1.8}] does an excellent job in estimating the temperature ratio $\Tr/\Tt$, as confirmed by simulations.\cite{LHMZ98,Z06} This is especially true for both small ($\beta=-0.9$) and large ($\be=0.9$) roughness. While the Maxwellian approximation overestimates the temperature ratio $\Tr/\Tt$ for $\be=-0.9$ and $\be=0$, it slightly underestimates this ratio for $\be=0.9$. It is interesting to note that, for nearly smooth spheres ($\beta=-0.9$), $\Tr/\Tt$ abruptly changes from small values for nearly elastic spheres ($\al\gtrsim 0.97$) to very large values for inelastic spheres ($\al\lesssim 0.95$). This effect becomes more and more dramatic as one approaches the smooth-sphere limit ($\beta\to -1$).\cite{SKG10}

As for the cumulants, Fig.\ \ref{fig3} shows some interesting features. In general, for large roughness ($\beta=0.9$) the magnitudes of the three cumulants are relatively small, meaning that the velocity distribution function is not far from a Maxwellian. This agrees with the almost indistinguishability between the Maxwellian and Sonine approximations observed in the bottom panel of Fig.\ \ref{fig2}. For medium roughness ($\be=0$) and large inelasticity ($\al\lesssim 0.9$), however, the cumulants reach relatively important values, especially in the case of $a_{11}$.  This trend is continued as roughness decreases ($\be=-0.9$), except in the case of $a_{02}$. The latter quantity takes a high maximum value at $\al\simeq 0.95$. This value is even higher than $1$, thus invalidating (at a quantitative level) the linearization method followed to estimate it. In any case, we expect that the Sonine method employed in this paper captures the main qualitative behavior of the cumulants for $\be=-0.9$ and $\al\simeq 0.95$. The peculiar change in the behavior of the cumulants when going from inelastic to nearly elastic spheres for small roughness ($\be=-0.9$) is correlated to the one observed in the case of the temperature ratio.

\begin{figure}
  \includegraphics[width=.95\columnwidth]{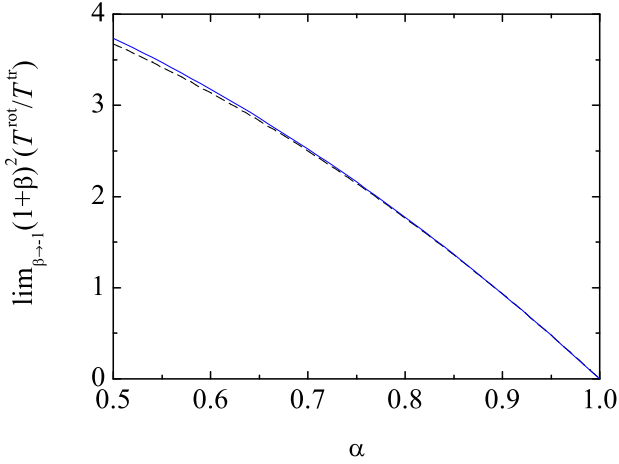}
\caption{(Color online) Plot of the HCS limit value $\lim_{\be\to -1}(1+\be)^2\Tr/\Tt$ vs the coefficient of normal restitution $\alpha$. The dashed and solid lines are the Maxwellian and Sonine approximations, respectively.} \label{fig4}
\end{figure}

\begin{figure}
  \includegraphics[width=.95\columnwidth]{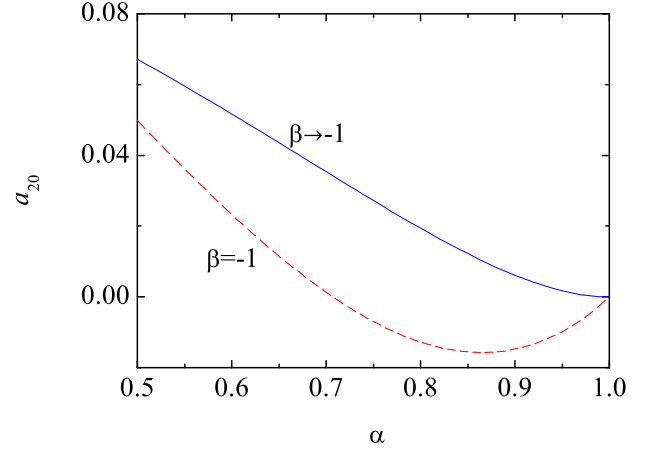}
\caption{(Color online) Plot of the HCS cumulant $a_{20}$ vs the coefficient of normal restitution $\alpha$ for $\be=-1$ (dashed line) and in the limit $\be\to -1$ (solid line).} \label{fig5}
\end{figure}

The smooth-sphere limit $\be\to -1$ (or, equivalently, $\bt\to 0$) deserves a separate treatment. According to  Eq.\ \eqref{1.8}, one can expect in that limit\cite{SKG10} the asymptotic behaviors
\beq
\theta\approx
\begin{cases}
\vartheta_1 \q\bt^{-2},&\al<1,\\
\vartheta_2 \bt,&\al=1,
\end{cases}
\label{4.6}
\eeq
with $\vartheta_1=\at(1-\at)$ and $\vartheta_2=1/(1-\q)$ in the Maxwellian approximation. Let us assume that $\alpha<1$. Thus, inserting $\theta=\vartheta_1\q\bt^{-2}$ into Eqs.\ \eqref{22}, \eqref{23}, and \eqref{28}--\eqref{30}, and taking the limit $\bt\to 0$, one gets
\beqa
\mu_{20}&=&4\sqrt{2\pi}\left[\at(1-\at)\left(1+\frac{3a_{20}}{16}  \right)\right.\nn
&&\left.-{\vartheta_1}\left(1-\frac{a_{20}}{16}+\frac{a_{11}}{4}\right)\right],
\label{4.7}
\eeqa
\beq
\mu_{02}=4\sqrt{2\pi}\frac{\bt}{\kappa}\left(1-\frac{a_{20}}{16}+\frac{a_{11}}{4}\right),
\label{4.8}
\eeq
\beqa
\mu_{40}&=&16\sqrt{2\pi}
\left\{\at^3(2-\at)+\frac{\at}{8}\left(11-19\at\right)\right.\nn
&&+\left[\at^3(2-\at)+\frac{\at}{120}\left(269-357\at\right)\right]\frac{15a_{20}}{16}\nn
&&-\frac{11 \vartheta_1}{8}\left(1+\frac{41a_{20}}{176 }+\frac{3a_{11}}{4}\right)+{\vartheta_1}\at(1-\at)\nn
&&\times\left(1+\frac{3a_{20}}{16}+\frac{3a_{11}}{4}\right)-{\vartheta_1^2}\left(1-\frac{a_{20}}{16}\right.\nn
&&\left.\left.+\frac{a_{11}}{2}+\frac{a_{02}}{2}\right)\right\},
\label{4.9}
\eeqa
\beqa
\mu_{22}&=&
3\sqrt{2\pi }\left[2\at(1-\at)
\left(1+\frac{3a_{20}}{16}+\frac{3a_{11}}{4}\right)\right.\nn
&&\left.-\frac{8\vartheta_1}{3}\left(1-\frac{a_{20}}{16}+\frac{a_{11}}{2}+\frac{5a_{02}}{8}\right)+\at a_{11}\right],\nn
\label{4.10}
\eeqa
\beq
\mu_{04}=20\sqrt{2\pi }\frac{\bt}{\kappa}\left(1-\frac{a_{20}}{16}+\frac{a_{11}}{2}+a_{02}\right).
\label{4.11}
\eeq
Since $\mu_{02}=\mathcal{O}(\bt)$, Eqs.\ \eqref{12}--\eqref{14} imply that, in the smooth-sphere limit,
\beq
\mu_{20}=0,\quad \mu_{40}=0,\quad \mu_{22}=0.
\label{4.12}
\eeq
Substitution of Eqs.\ \eqref{4.8} and \eqref{4.11} into Eq.\ \eqref{15bb}, and neglecting nonlinear terms, gives $a_{11}=0$. Next, from Eqs.\ \eqref{4.7} and \eqref{4.10}, together with $\mu_{20}=\mu_{22}=0$, we obtain
\beq
a_{20}=16\frac{\vartheta_1-\at(1-\at)}{\vartheta_1+3\at(1-\at)},
\label{4.13}
\eeq
\beq
a_{02}=-\frac{8}{5}\frac{\at(1-\at)}{\vartheta_1+3\at(1-\at)}.
\label{4.14}
\eeq
Finally, from Eqs.\ \eqref{4.9}, \eqref{4.13}, and \eqref{4.14}, together with the condition $\mu_{40}=0$, we get a closed quadratic equation for $\vartheta_1$,
\beqa
[65&-&8\at(1-\at)]\vartheta_1^2-40\at[9-12\at+4\at^2(2-\at)]\vartheta_1\nn
&
+&5\at^2(1-\at)[59-75\at
+24\at^2(2-\at)]=0.
\label{4.15}
\eeqa

Figure \ref{fig4} shows $\lim_{\be\to -1}(1+\be)^2\Tr/\Tt=[4(1+\q)^2/\q]\vartheta_1$, where $\vartheta_1$ is the physical solution of Eq.\ \eqref{4.15}, as a function of $\alpha$. We observe again a very good agreement between the Maxwellian and the Sonine approximations.

Insertion of the solution of Eq.\ \eqref{4.15}  into Eqs.\ \eqref{4.13} and \eqref{4.14} gives $a_{20}$ and $a_{02}$, respectively. In the elastic limit ($\al\to 1$) Eq.\ \eqref{4.15} yields $\vartheta_1\to 1-\at$, what implies $a_{20}\to 0$ and $a_{02}\to -\frac{2}{5}$. In fact, $\vartheta_1\approx \at(1-\at)$ even in the inelastic case, so that  $a_{02}\approx -\frac{2}{5}$ for all $\al$.

The cumulant $\lim_{\be\to -1}a_{20}$ is plotted in Figure \ref{fig5}. For comparison, this figure also includes the curve representing $a_{20}$ in the pure smooth case ($\be=-1$ from the very beginning). In the latter case, the translational and rotational degrees of freedom are absolutely decoupled and, in addition, each particle keeps its initial angular velocity, so the arbitrary initial distribution of angular velocities does not change with time. This implies that only Eq.\ \eqref{13} keeps being meaningful if $\be=-1$. Inserting Eqs.\ \eqref{22} and \eqref{28} with $\bt=0$ into Eq.\ \eqref{13} one gets\cite{MS00,SM09}
\beqa
a_{20}&=&-16(1-\at)\frac{1 - 8 \at(1-\at)}{63 - 23 \at - 8 \at^2(2-\at)}\nn
&=&16(1-\al)\frac{1 - 2\al^2}{97 - 33 \al - 2 \al^2(1-\al)}.
\label{4.16}
\eeqa
We observe that $a_{20}$ at $\beta=-1$ differs from $\lim_{\be\to -1}a_{20}$. This singular effect of $a_{20}$ is analogous to the one observed {for the ratio $\langle (\mathbf{c}\cdot\mathbf{w})^2\rangle/\langle c^2w^2\rangle$,\cite{BPKZ07,KBPZ09} as well as} in the case of the translational/translational temperature ratio in mixtures.\cite{SKG10,S10a} The explanation of this interesting phenomenon is similar in {all these} situations. While for smooth particles ($\beta=-1$) the rotational temperature is totally isolated from the translational one (so that the dotted arrows  in Fig.\ \ref{fig1} disappear),  in the case of  quasi-smooth particles ($\beta\gtrsim -1$) a weak channel of energy transfer exists between the rotational and translational degrees of freedom and also the rotational temperature is subject to a weak cooling. Even though $\be$ might be very close to $-1$, the transfer of energy  eventually becomes activated when the rotational temperature is sufficiently larger than the translational one. In other words, even if the dotted arrows in Fig.\ \ref{fig1} are very weak, the great disparity between $\Tr$ and $\Tt$ triggers the flux of energy from the rotational toward the translational degrees of freedom, thus significantly modifying the cumulant $a_{20}$ with respect to the case of strict smooth spheres.

\section{Application to the white-noise thermostat\label{sec5}}
As a second application, we consider now a homogeneous granular gas subject to a stochastic thermostat force with properties of a Gaussian white noise.\cite{WM96,W96,SBCM98,vNE98,MS00} The corresponding Boltzmann equation reads\cite{vNE98}
\beq
\partial_t f+\cc\cdot \nabla f-\frac{\chi_0^2}{2}\left(\frac{\partial}{\partial \cc}\right)^2f={J[\cc,\ww|f]},
\label{5.1}
\eeq
where $\chi_0^2$ is a measure of the strength of the stochastic force. This force acts as a ``thermostat'' that injects energy to the system, thus compensating for the collisional energy loss until a steady state is eventually reached. The evolution equations for the temperatures are
\beq
\partial_t\Tt-{m\chi_0^2}=-\zt \Tt,\quad \partial_t\Tr=-\zr \Tr,
\label{5.2}
\eeq
\beq
\partial_t T-\frac{m\chi_0^2}{2}=-\zeta  T.
\label{5.3}
\eeq

In terms of the dimensionless  variables defined by Eqs.\ \eqref{7a} and \eqref{7b}, Eq.\ \eqref{5.1} becomes
\beqa
\partial_s\phi&&+\frac{\mu_{20}-3\Gamma}{3}\frac{\partial}{\partial\mathbf{c}}\cdot\left(\mathbf{c}\phi\right)
+\frac{\mu_{02}}{3}\frac{\partial}{\partial\mathbf{w}}\cdot\left(\mathbf{w}\phi\right)\nn
&&-
\frac{\Gamma}{2}\left(\frac{\partial}{\partial \mathbf{c}}\right)^2\phi= {J^*[\mathbf{c},\mathbf{w}|\phi]},
\label{5.4}
\eeqa
where
\beq
\Gamma\equiv \frac{\chi_0^2}{n\sigma^2(2\Tt/m)^{3/2}}.
\label{5.5}
\eeq
After taking moments in Eq.\ \eqref{5.4} one obtains
\beqa
{
-\partial_s \langle c^pw^q\rangle}&&{+\frac{1}{3}\left[p\left(\mu_{20}-3\Gamma\right)+q\mu_{02}\right]\langle c^pw^q\rangle
}
\nn
&&
{
+\frac{\Gamma}{2}p(p+1)
\langle c^{p-2}w^q\rangle=\mu_{pq}.
}
\label{5.7}
\eeqa

In the steady state, Eq.\ \eqref{5.2} implies that $\zt\Tt=m\chi_0^2$ and $\zr=0$. Equivalently,
\beq
\mu_{20}=3\Gamma,
\label{5.8}
\eeq
\beq
\mu_{02}=0.
\label{5.9}
\eeq
As a consequence,  {Eq.\  \eqref{5.7} yields}, in the steady-state,
\beq
{5\mu_{20}=\mu_{40},}
\label{5.10}
\eeq
\beq
\frac{3}{2}\mu_{20}=\mu_{22},
\label{5.11}
\eeq
\beq
\mu_{04}=0.
\label{5.12}
\eeq

\begin{figure}
  \includegraphics[width=.95\columnwidth]{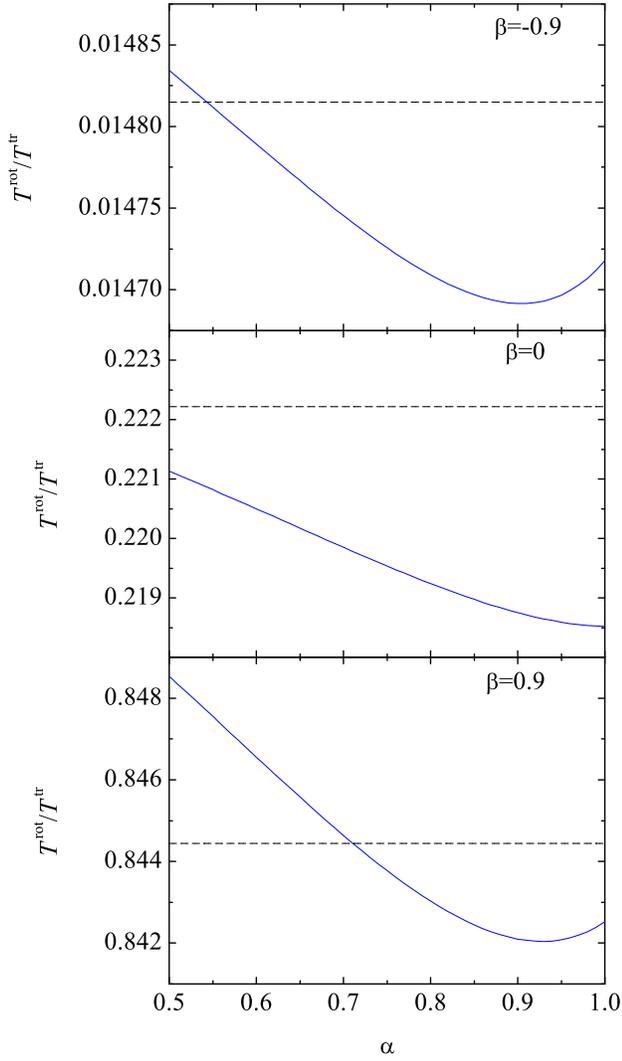}
\caption{(Color online) Plot of the WNS temperature ratio $\Tr/\Tt$ vs the coefficient of normal restitution $\alpha$ for $\beta=-0.9$ (top panel), $\beta=0$ (middle panel), and $\beta=0.9$ (bottom panel).
The dashed and solid lines are the Maxwellian and Sonine approximations, respectively.} \label{fig6}
\end{figure}

\begin{figure}
 \includegraphics[width=.95\columnwidth]{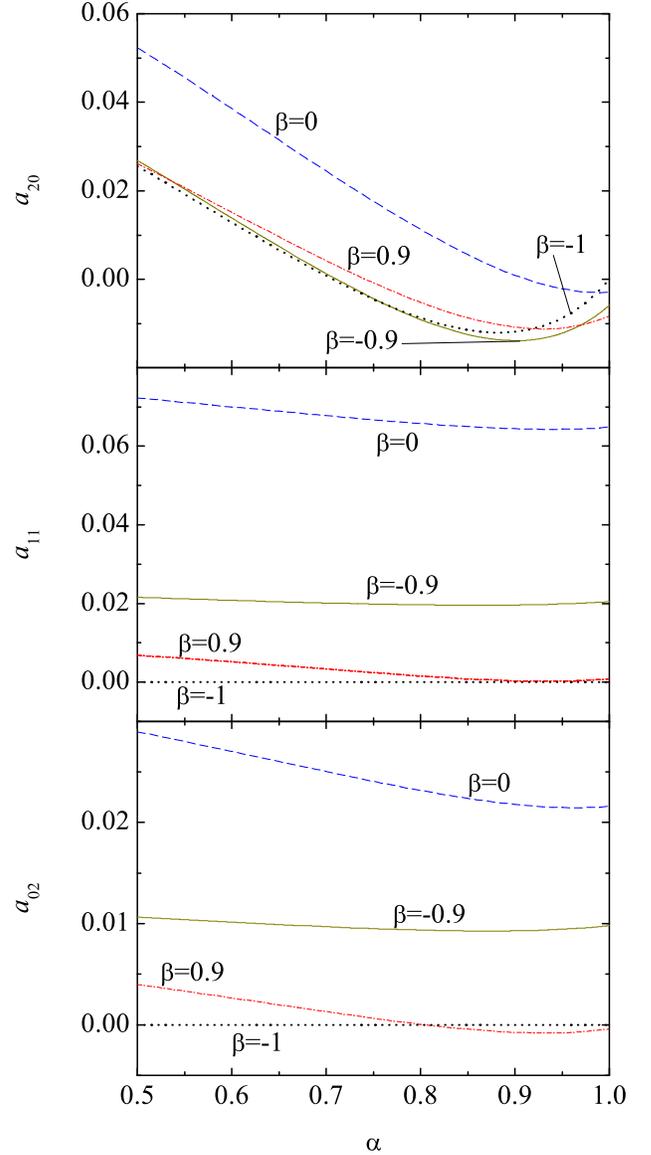}
\caption{(Color online) Plot of the WNS cumulants $a_{20}$ (top panel), $a_{11}$ (middle panel), and $a_{02}$ (bottom panel) vs the coefficient of normal restitution  $\al$ for $\beta=-1$ (dotted lines), $\beta=-0.9$ (solid lines), $\beta=0$ (dashed lines), and $\beta=0.9$ (dash-dotted lines).} \label{fig7}
\end{figure}

\begin{figure}
  \includegraphics[width=.95\columnwidth]{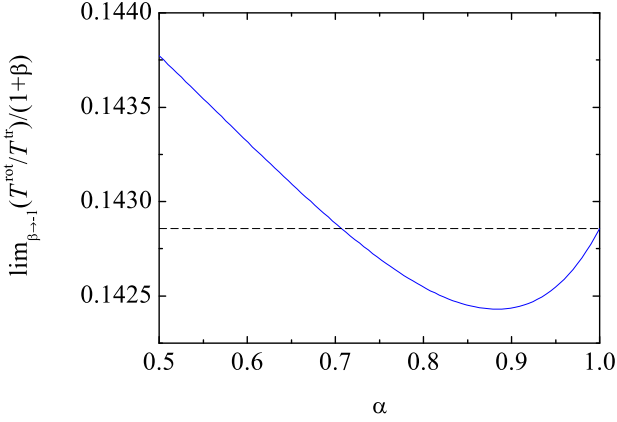}
\caption{(Color online) Plot of the WNS limit value $\lim_{\be\to -1}(\Tr/\Tt)/(1+\be)$ vs the coefficient of normal restitution $\alpha$. The dashed and solid lines are the Maxwellian and Sonine approximations, respectively.} \label{fig8}
\end{figure}

{Analogously to the HCS case,} taking into account Eqs.\ \eqref{22}, \eqref{23}, and \eqref{28} in the steady-state conditions \eqref{5.9}--\eqref{5.11} one obtains the cumulants $a_{20}$, $a_{11}$, and $a_{02}$ as functions of $\theta$. Next, Eqs.\ \eqref{30} and \eqref{5.12} provide a closed cubic equation for $\theta$.

Figure \ref{fig6} shows the temperature ratio as a function of $\alpha$ for $\beta=-0.9$, $0$, and $0.9$. In the Maxwellian approximation [cf. Eq.\ \eqref{1.10}] $\theta$ is independent of $\al$. On the other hand, we observe that the Sonine approximation predicts a very \emph{weak} dependence on $\alpha$ {(note the vertical scales)}. Otherwise, the maximum deviation between both approximations in the range $0.5\leq \al\leq 1$ is smaller than $1$\%, $2$\%, and $0.5$\% for $\beta=-0.9$, $0$, and $0.9$, respectively. It is clearly apparent that $\Tr/\Tt$ increases with increasing roughness,  ranging from $0$ in the smooth-sphere limit ($\be=-1$) to about $1$ in the opposite limit $\be\to 1$. In the latter limit the Maxwellian approximation gives $\Tr/\Tt=1$ for all $\al$, while the Sonine approximation gives $\Tr/\Tt\simeq 1.01$ for $\al=0$.

The cumulants are plotted in Fig.\ \ref{fig7}. Their magnitudes are much smaller than in the HCS case (compare with Fig.\ \ref{fig3}). Apart from that, they are more significant for medium roughness ($\be=0$) than for large ($\be=1$) or small ($\be=-0.9$) roughness.

Again, it is worth analyzing separately the smooth-sphere limit $\be\to -1$. {}From Eq.\ \eqref{1.10} we can expect
\beq
\theta\approx \vartheta \bt,
\label{5.13}
\eeq
where $\vartheta=1$ in the Maxwellian approximation.
Therefore,  in the limit $\bt\to 0$ Eqs.\ \eqref{22}, \eqref{23}, and \eqref{28}--\eqref{30} become
\beq
\mu_{20}=4\sqrt{2\pi}\at(1-\at)\left(1+\frac{3a_{20}}{16}\right),
\label{5.14}
\eeq
\beq
\mu_{02}=4\sqrt{2\pi}\frac{\bt}{\kappa}\left[1-\frac{a_{20}}{16}+\frac{a_{11}}{4}-\frac{1}{\vartheta}\left(1+\frac{3a_{20}}{16}\right)\right],
\label{5.15}
\eeq
\beqa
\mu_{40}&=&16\sqrt{2\pi}\at
\left\{\at^2(2-\at)+\frac{1}{8}\left(11-19\at\right)\right.\nn
&&\left.+\left[\at^2(2-\at)+\frac{1}{120}\left(269-357\at\right)\right]\frac{15a_{20}}{16}\right\},\nn
\label{5.16}
\eeqa
\beq
\mu_{22}=
3\sqrt{2\pi }\at
\left[2(1-\at)\left(1+\frac{3a_{20}}{16}+\frac{3a_{11}}{4}\right)+ a_{11}\right],
\label{5.17}
\eeq
\beqa
\mu_{04}&=&20\sqrt{2\pi }\frac{\bt}{\kappa}\left[1-\frac{a_{20}}{16}+\frac{a_{11}}{2}+a_{02}
\right.\nn
&&\left.-\frac{1}{\vartheta}\left(1+\frac{3a_{20}}{16}+\frac{3a_{11}}{4}\right)
\right].
\label{5.18}
\eeqa
Equations \eqref{5.14} and \eqref{5.16}  include only the parameter $a_{20}$ and so they are exactly the same as those obtained in the case of pure smooth spheres.\cite{vNE98} Therefore, application of the steady-state condition \eqref{5.10} gives
\beqa
a_{20}&=&-16(1-\at)\frac{1 - 8 \at(1-\at)}{239 -327  \at +120 \at^2(2-\at)}\nn
&=&16(1-\al)\frac{1 - 2\al^2}{241 - 177 \al +30 \al^2(1-\al)}.
\label{5.19}
\eeqa
Thus, in contrast to the HCS case, the cumulant $a_{20}$ in the WNS coincides in the limit $\be\to -1$ with that at $\be=-1$. The Sonine approximation in that limit, Eq.\ \eqref{5.19}, is also plotted in Fig.\ \ref{fig7}, where we observe that the curve is close to the one for $\be=-0.9$, especially for large inelasticity.

Application of Eqs.\ \eqref{5.14} and \eqref{5.17} in Eq.\ \eqref{5.11} implies $a_{11}=0$. Next, from Eqs.\ \eqref{5.9}, \eqref{5.15}, and \eqref{5.19} we get
\beqa
\vartheta&=&\frac{1+\frac{3}{16}a_{20}}{1-\frac{1}{16}a_{20}}\nn
&=&\frac{61-45\al+ 6\al^2(1-\al)}{4[15 - 11 \al +2 \al^2(1-\al)]}.
\label{5.20}
\eeqa
Finally, Eqs.\ \eqref{5.12} and \eqref{5.18} imply $a_{02}=0$. Therefore, in the limit $\be\to -1$ the translational distribution function is non-Maxwellian, as measured by $a_{20}\neq 0$, but the rotational distribution tends to a Maxwellian ($a_{11}=a_{02}=0$) with a temperature $\Tr$ much smaller than $\Tt$. Note that if the granular gas is made of strict smooth spheres ($\be=-1$) the rotational distribution function (and hence the temperature $\Tr$) is not uniquely defined since it preserves its initial form. The parameter $\lim_{\be\to -1}(\Tr/\Tt)/(1+\be)=[\q/2(1+\q)]\vartheta$
is plotted in Fig.\ \ref{fig8}. Again, the relative difference between the Maxwellian and the Sonine predictions is quite small (less than $0.7$\% in the range $0.5\leq\al\leq 1$).

\section{Discussion and concluding remarks\label{sec6}}

The primary goal of this paper has been the derivation, within  the Sonine approximation given by Eq.\ \eqref{20}, of the second- and fourth-degree collisional moments [cf.\ {Eq.\ \eqref{10b}}] in a granular gas made of inelastic rough hard spheres. The results are given by {Eqs.\ \eqref{22}, \eqref{23},} and \eqref{28}--\eqref{30}. In particular, the second-degree collisional moments $\mu_{20}$ and $\mu_{02}$ are not but dimensionless versions of the  collisional rates of change $\zt$, $\zr$, and $\zeta$ associated with the temperatures $\Tt$, $\Tr$, and $T$, respectively [cf.\ Eqs.\ \eqref{3.6} and \eqref{3.9bis}]. {}From that point of view, it is also worth emphasizing  that Eqs.\ \eqref{3.2}--\eqref{3.4} are exactly derived from the Boltzmann equation without any further assumption, so that they are not restricted to any Sonine approximation.

Our results represent extensions of some previously derived results. On the one hand, Eqs.\ \eqref{22}--\eqref{3.9} are Sonine extensions of  those obtained in the Maxwellian approximation defined by Eq.\  \eqref{1.11}.\cite{Z06,SKG10} On the other hand, Eqs.\ \eqref{22} and \eqref{28} are extensions to rough spheres of previous Sonine derivations for smooth spheres.\cite{vNE98}

Since the Boltzmann collision operator [cf.\ Eq.\ \eqref{2.2}] is local in time and space, Eqs.\ {\eqref{22}--\eqref{30}} keep being applicable to inhomogeneous and unsteady states. They also hold in the context of the Enskog equation for homogeneous states, except that the collision frequency \eqref{1.5} must be multiplied by the pair correlation function at contact, $g(\sigma)$. Apart from that, it is important to bear in mind that some restrictions apply to the Sonine approximation \eqref{20}. First, it has been assumed that the mean angular velocity vanishes, i.e., $\langle \ww \rangle=\mathbf{0}$. This restriction is easy to circumvent by replacing $\ww\to\ww-\langle \ww \rangle$ and $\Tr\to{\Trb}$ [see discussion below Eq.\ \eqref{3.1}] in Eqs.\ \eqref{1.11}, \eqref{7a}, and \eqref{7b}. These changes would affect the collision rules in Eqs.\ {\eqref{25}--\eqref{27} by the changes $\theta\to \Trb/\Tt$ and $\mathbf{w}_1+\mathbf{w}_2\to \mathbf{w}_1+\mathbf{w}_2+2\langle\ww\rangle/({2\Trb/I})^{1/2}$}, given that the angular velocity is not a a conserved quantity. For the expressions of $\zt$ and $\zr$ in the Maxwellian approximation with $\langle \ww \rangle\neq\mathbf{0}$, the reader is referred to Refs.\ \onlinecite {SKG10,S10b}.

As a second restriction, notice that Eq.\ \eqref{20} may be less useful in strongly anisotropic states where the dependence of $\phi(\mathbf{c},\mathbf{w})$ on the six velocity components is not exhausted by the three scalar quantities {$c^2$, $w^2$, and $(\mathbf{c}\cdot\mathbf{w})^2$.}
Finally, while Eq.\ \eqref{20} treats the {two} second-degree moments as independent quantities, it does not do so with the \emph{a priori} {four} independent fourth-degree moments. Instead, Eq.\ \eqref{20} assumes that the {moment $\langle (\mathbf{c}\cdot\mathbf{w})^2\rangle$ is enslaved to  $\langle c^2w^2\rangle$  by Eq.\ \eqref{25.2}. Therefore, the study of the orientational correlation between the translational and angular velocities has not been included in our scheme.}

We have applied Eqs.\ {\eqref{22}--\eqref{30}} to two paradigmatic homogeneous and isotropic situations: the similarity solution of the HCS and the steady-state solution of the WNS. {In both cases we have found that} the Maxwellian approximation for the temperature ratio $\Tr/\Tt$, being much simpler than the corresponding Sonine approximation, does a very good job, especially for large or small roughness.
On the other hand, departures of the velocity distribution function from the Maxwellian, as measured by the cumulants $a_{20}$, $a_{11}$, and $a_{02}$, cannot be ignored. This is especially true in the case of the HCS for medium and small roughness. In fact, in the quasi-smooth limit $\be\to -1$ the HCS results differ markedly from those obtained in the case of pure smooth spheres ($\be=-1$). This interesting singular behavior is directly related to the unsteady character of the HCS and thus it is absent in the steady WNS.

We expect that this work can contribute to our understanding of the subtle interplay between roughness and inelasticity in granular gases and how the former feature modifies the properties of inelastic smooth spheres. We plan to assess the qualitative and quantitative results derived here by comparison with computer simulations for several situations of physical interest.

\acknowledgments
This paper is dedicated to the memory of Carlo Cercignani.
The work of A.S. has been supported by the
Ministerio de  Ciencia e Innovaci\'on (Spain) through Grant No. FIS2010-16587 (partially financed by FEDER funds).
The work of G.M.K. and M.S. has been supported by the Conselho Nacional de Desenvolvimento Cient\'{\i}fico e
Tecnol\'ogico (Brazil).


\begin{thebibliography}{99}

\bibitem{CCG00}
J. A. Carrillo, C. Cercignani,  and I. M. Gamba, ``Steady states of a Boltzmann equation for driven granular media,''
Phys. Rev. E \textbf{62} 7700 (2000).

\bibitem{C01}
 C. Cercignani, ``Shear flow of a granular material,''
J. Stat. Phys.  \textbf{102}, 1407 (2001).

\bibitem{CIS01}
 C. Cercignani, R. Illner, and C. Stoica, ``On diffusive equilibria in generalized kinetic theory,''
J. Stat. Phys.  \textbf{105}, 337 (2001).

\bibitem{C02}
 C. Cercignani, ``The Boltzmann equation approach to the shear flow of a granular material,'' Phil. Trans. R. Soc. Lond. A \textbf{360}, 407 (2002).

\bibitem{BC02}
A. V. Bobylev and C. Cercignani, ``Moment equations for a granular material in a thermal bath,''
J. Stat. Phys.  \textbf{106}, 547 (2002).

\bibitem{BC03}
A. V. Bobylev and C. Cercignani, ``Self-similar asymptotics for the Boltzmann equation with inelastic and elastic interactions,''
J. Stat. Phys.  \textbf{110}, 333 (2003).

\bibitem{BCG03}
A. V. Bobylev, C. Cercignani, and G. Toscani, ``Proof of an asymptotic property of self-similar solutions of the Boltzmann equation for granular materials,''
J. Stat. Phys.  \textbf{111}, 403 (2003).

\bibitem{BCG08}
A. V. Bobylev, C. Cercignani, and I. M. Gamba, ``Generalized Kinetic Maxwell Type Models of Granular Gases,'' in \emph{Mathematical Models of Granular Matter}, edited by G. Capriz, P. Giovine, and P. M. Mariano, Lecture Notes in Mathematics, Vol.\ 1937 (Springer, Berlin, 2008), pp.\ 23--57

\bibitem{BCG09}
A. V. Bobylev, C. Cercignani, and I. M. Gamba, ``On the Self-Similar Asymptotics for Generalized Nonlinear Kinetic Maxwell Models,'' Comm. Math. Phys. \textbf{291}, 594 (2009).


\bibitem{G03}
I. Goldhirsch,  ``Rapid granular flows,'' Annu.\ Rev.\ Fluid Mech. \textbf{35}, 267 (2003).



\bibitem{MSS04}
S. J. Moon, J. B. Swift, and H. L. Swinney, ``Steady-state velocity distributions of an oscillates granular gas.''  Phys. Rev. E \textbf{69}, 011301 (2004).

\bibitem{JR85}
J. T. Jenkins and M. W. Richman, ``Kinetic theory for plane flows of a dense gas of identical, rough, inelastic, circular disks,''  Phys.\ Fluids \textbf{28}, 3485 (1985).

\bibitem{LS87}
C. K. K. Lun and S. B. Savage, ``A Simple Kinetic Theory for Granular Flow of Rough, Inelastic, Spherical Particles,'' J.\ Appl.\ Mech. \textbf{54}, 47 (1987).

\bibitem{C89}
C. S. Campbell, ``The stress tensor for simple shear flows of a granular material,'' J.\ Fluid Mech. \textbf{203}, 449 (1989).

\bibitem{L91}
C. K. K. Lun,  ``Kinetic theory for granular flow of dense, slightly inelastic, slightly rough spheres,'' J.\ Fluid Mech. \textbf{233}, 539 (1991).

\bibitem{LN94}
{C. K. K. Lun and A. A. Bent,  ``Numerical simulation of inelastic spheres in simple shear flow,'' J.\ Fluid Mech. \textbf{258}, 335 (1994).}

\bibitem{L96}
C. K. K. Lun,  ``Numerical simulation of inelastic spheres in simple shear flow,'' Phys.\ Fluids \textbf{8}, 2868 (1996).

\bibitem{ZVPSH98}
P. Zamankhan, H. V. Tafreshi, W. Polashenski, P. Sarkomaa, and C. L. Hyndman,  ``Shear induced diffusive mixing in simulations of dense Couette flow of rough, inelastic hard spheres,'' J.\ Chem.\ Phys. \textbf{109}, 4487 (1998).

\bibitem{JZ02}
J. T. Jenkins and C. Zhang,   ``Kinetic theory for identical, frictional, nearly elastic spheres,'' Phys.\ Fluids \textbf{14}, 1228 (2002).

\bibitem{PZMZ02}
W. Polashenski,  P. Zamankhan, S. M\"akiharju, and P. Zamankhan,  ``Fine structures in sheared granular flows,'' Phys. Rev. E \textbf{66}, 021303 (2002).

\bibitem{GNB05}
I. Goldhirsch, S. H. Noskowicz, and O. Bar-Lev,  ``Nearly Smooth Granular Gases,'' Phys. Rev. Lett. \textbf{95}, 068002 (2005).


\bibitem{GS95}
A. Goldshtein and M. Shapiro,  ``Mechanics of collisional motion of granular materials. Part 1. General hydrodynamic equations,'' J.\ Fluid Mech. \textbf{282}, 75 (1995).



\bibitem{HZ98}
M. Huthmann,  and A. Zippelius,  ``Dynamics of inelastically colliding rough spheres: Relaxation of translational and rotational energy,'' Phys. Rev. E \textbf{56}, R6275 (1998).

\bibitem{ML98}
S. McNamara and S. Luding,   ``Energy non-equipartition in systems of inelastic, rough spheres,'' Phys. Rev. E \textbf{58}, 2247 (1998).

\bibitem{LHMZ98}
S. Luding, M. Huthmann, S. McNamara, and A. Zippelius, ``Homogeneous cooling of rough, dissipative particles: Theory and simulations,'' Phys. Rev. E \textbf{58}, 3416 (1998).

\bibitem{HHZ00}
O. Herbst, M. Huthmann, and A. Zippelius,  ``Dynamics of inelastically colliding spheres with Coulomb friction: Relaxation of translational and rotational energy,'' Gran.\ Matt. \textbf{2}, 211 (2000).

\bibitem{AMZ01}
T. Aspelmeier, M. Huthmann, and A. Zippelius, ``Free Cooling of Particles with Rotational Degrees of Freedom,'' in \emph{Granular Gases}, edited by T. P\"oschel and S. Luding (Springer, Berlin, 2001), pp.\ 31--58.

\bibitem{CLH02}
R. Cafiero, S. Luding, and H. J. Herrmann,  ``Rotationally driven gas of inelastic rough spheres,'' Europhys.\ Lett. \textbf{60}, 854 (2002).

\bibitem{Z06}
A. Zippelius,  ``Granular gases,'' Physica A \textbf{369}, 143 (2006).

\bibitem{L95}
S. Luding,  Phys. Rev. E ``Granular materials under vibration: Simulations of rotating spheres,'' \textbf{52}, 4442 (1995).

\bibitem{MHN02}
N. Mitarai, H. Hayakawa, and H. Nakanishi,  ``Collisional Granular Flow as a Micropolar Fluid,'' Phys. Rev. Lett. \textbf{88}, 174301 (2002).

\bibitem{BPKZ07}
N. V. Brilliantov, T. P\"oschel, W. T. Kranz, and A. Zippelius,  ``Translations and Rotations Are Correlated in Granular Gases,'' Phys. Rev. Lett. \textbf{98}, 128001 (2007).

\bibitem{KBPZ09}
{W. T. Kranz, N. V. Brilliantov, T. P\"oschel,  and A. Zippelius,  ``Correlation of spin and velocity in the homogeneous
cooling state of a granular gas of rough particles,'' Eur. Phys. J. Spec. Top. \textbf{179}, 91 (2009).}

\bibitem{SKG10}
A. Santos, G. M. Kremer, and V. Garz\'o, ``Energy production rates in fluid mixtures of inelastic rough hard spheres,'' Prog. Theor. Phys. Suppl. \textbf{184}, 31 (2010).


\bibitem{S10a}
A. Santos, ``Homogeneous Free Cooling State in Binary Granular Fluids of Inelastic Rough Hard Spheres,''
{in Rarefied Gas Dynamics: Proceedings of the 27th International Symposium on Rarefied Gas Dynamics, D. A. Levin, ed. (AIP Conference Proceedings, Melville, NY, 2011), in press;}
preprint arXiv:1007.0701.

\bibitem{S10b}
A. Santos, ``A Bhatnagar--Gross--Krook-like Model Kinetic Equation for a Granular Gas of Inelastic Rough Hard Spheres,''
{in Rarefied Gas Dynamics: Proceedings of the 27th International Symposium on Rarefied Gas Dynamics, D. A. Levin, ed. (AIP Conference Proceedings, Melville, NY, 2011), in press;} preprint arXiv:1007.0700.


\bibitem{UKAZ09}
H. Uecker, W. T. Kranz, T. Aspelmeier, and A. Zippelius, ``Partitioning of energy in highly polydisperse granular gases,'' {Phys. Rev. E} \textbf{80}, 041303 (2009).

\bibitem{WM96}
D. R. M. Williams and F. C. MacKintosh, ``Driven granular media in one dimension: correlations and equation of state,'' Phys. Rev. E \textbf{54}, R9 (1996).

\bibitem{W96}
D. R. M. Williams, ``Driven granular media and dissipative gases: correlations and liquid-gas phase transitions,'' Physica A \textbf{233}, 718 (1996).



\bibitem{SBCM98}
M. R. Swift, M. Boamf\v{a}, S. J. Cornell, and A. Maritan, ``Scale invariant correlations in a driven dissipative gas,'' Phys. Rev. Lett. \textbf{80}, 4410 (1998).

\bibitem{vNE98}
T. P. C. van Noije and M. H. Ernst, ``Velocity distributions in homogeneous granular fluids: the free and the heated case,'' Gran. Matt. \textbf{1}, 57 (1998).

\bibitem{MS00}
J. M. Montanero and A. Santos, ``Computer simulation of uniformly heated granular fluids,'' Gran. Matt. \textbf{2}, 53 (2000).


\bibitem{SM09}
A. Santos and J. M. Montanero,  ``The second and third Sonine coefficients of a freely cooling granular gas revisited,'' Gran. Matt. \textbf{11}, 157 (2009).

\bibitem{BDS97}
J. J. Brey, J. W. Dufty, and A. Santos, ``Dissipative dynamics for hard spheres,'' J. Stat. Phys. \textbf{87}, 1051 (1997).













\end{thebibliography}
\end{document}